\input{aipcheck}

\documentclass[
    ,final            
  ]
  {aipproc}

\layoutstyle{6x9}

\begin{document}

\title{Isospin dependence of the $\eta'$ meson production in 
       nucleon--nucleon collisions}

\classification{\texttt{  13.85.Ni  }}

\author{J.~Przerwa}{
  address={Nuclear Physics Department, Jagellonian University, 30-059 Cracow, Poland}
}

\author{H.-H.~Adam}{
  address={Institut f{\"u}r Kernphysik, Universist\"at M\"unster, 48419 M\"unster, Germany}
}

\author{A.~Budzanowski}{
  address={Institute of Nuclear Physics, 31-342 Cracow, Poland}
}

\author{R.Czy{\.z}ykiewicz}{
   address={Institut f\"ur Kernpfysik, Forschungszentrum J\"ulich, 52425 J\"ulich, Germany}
  ,altaddress={Nuclear Physics Department, Jagellonian University, 30-059 Cracow, Poland}
}

\author{D.~Grzonka}{
   address={Institut f\"ur Kernpfysik, Forschungszentrum J\"ulich, 52425 J\"ulich, Germany}
}

\author{M.~Janusz}{
  address={Nuclear Physics Department, Jagellonian University, 30-059 Cracow, Poland}
}

\author{L.~Jarczyk}{
  address={Nuclear Physics Department, Jagellonian University, 30-059 Cracow, Poland}
}

\author{B.~Kamys}{
  address={Nuclear Physics Department, Jagellonian University, 30-059 Cracow, Poland}
}

\author{A.~Khoukaz}{
  address={Institut f{\"u}r Kernphysik, Universist\"at M\"unster, 48419 M\"unster, Germany}
}

\author{K.~Kilian}{
   address={Institut f\"ur Kernpfysik, Forschungszentrum J\"ulich, 52425 J\"ulich, Germany}
}

\author{P.~Klaja}{
  address={Nuclear Physics Department, Jagellonian University, 30-059 Cracow, Poland}
}

\author{P.~Moskal}{
  address={Nuclear Physics Department, Jagellonian University, 30-059 Cracow, Poland}
  ,altaddress={Institut f\"ur Kernpfysik, Forschungszentrum J\"ulich, 52425 J\"ulich, Germany}
}

\author{W.~Oelert}{
   address={Institut f\"ur Kernpfysik, Forschungszentrum J\"ulich, 52425 J\"ulich, Germany}
}

\author{C.~Piskor--Ignatowicz}{
  address={Nuclear Physics Department, Jagellonian University, 30-059 Cracow, Poland}
}

\author{J.~Ritman}{
  address={Institut f\"ur Kernpfysik, Forschungszentrum J\"ulich, 52425 J\"ulich, Germany}
}

\author{T.~Ro\.zek}{
    address={Institut f\"ur Kernpfysik, Forschungszentrum J\"ulich, 52425 J\"ulich, Germany}
   ,altaddress={Institute of Physics, University of Silesia, 40-007 Katowice, Poland}
}

\author{T.~Sefzick}{
  address={Institut f\"ur Kernpfysik, Forschungszentrum J\"ulich, 52425 J\"ulich, Germany}
}

\author{M.~Siemaszko}{
  address={Institute of Physics, University of Silesia, 40-007 Katowice, Poland}
}

\author{J.~Smyrski}{
  address={Nuclear Physics Department, Jagellonian University, 30-059 Cracow, Poland}
}

\author{A.T\"aschner}{
   address={Institut f{\"u}r Kernphysik, Universist\"at M\"unster, 48419 M\"unster, Germany}
}

\author{J.~Wessels}{
    address={Institut f{\"u}r Kernphysik, Universist\"at M\"unster, 48419 M\"unster, Germany}
}

\author{P.~Winter}{
  address={Institut f\"ur Kernpfysik, Forschungszentrum J\"ulich, 52425 J\"ulich, Germany}
}

\author{M.~Wolke}{
  address={Institut f\"ur Kernpfysik, Forschungszentrum J\"ulich, 52425 J\"ulich, Germany}
}

\author{P.~W\"ustner}{
  address={ZEL Forschungszentrum J\"ulich, 52425 J\"ulich, Germany}
}
\author{W.~Zipper}{
  address={Institute of Physics, University of Silesia, 40-007 Katowice, Poland}
}

\begin{abstract}
According to the quark model, the masses of $\eta$ and $\eta'$ mesons
should be almost equal. However, the empirical values of these masses
differ by more than the factor of two. Similarly, though the almost the same quark-antiquark 
content, the total cross section for the creation of these mesons close to the kinematical
thresholds in the $pp \to ppX$ reaction differs significantly.
Using the COSY-11 detection setup we intend to determine whether this difference
will also be so significant in the case of the production of these mesons
in the proton-neutron scattering. Additionally, the comparison of the $ pp \to pp \eta'$
and $pn \to pn \eta'$ total cross sections will allow to learn about the production
of the $\eta'$ meson in the channels of isospin I = 0 and I = 1 and to investigate
aspects of the gluonium component of the $\eta'$ meson.\\

\end{abstract}

\maketitle


\section{Introduction}
Despite the fact that the $\eta'$ meson was observed fourty years ago,
its structure is still not known.
According to the quark model, $\eta$ and $\eta'$ mesons
can be described as the mixture of the singlet
$ (\eta_1 = {1 \over {\sqrt 3}} (u\bar u + d\bar d + s\bar s))$                                          
and octet $ (\eta_8 = {1 \over {\sqrt 6}} (u\bar u + d\bar d - 2 s\bar s))$
states of the SU(3) - flavour pseudoscalar meson nonet. Within the one mixing angle scheme,
a small mixing angle
$(\Theta = - 15.5^{0})$ implies that the masses of $\eta$ and $\eta'$ mesons
should be almost equal. However, the empirical values of these masses
differ by more than the factor of two, and the mass of the $\eta'$
does not fit utterly to the SU(3) scheme.\\
More surprisingly the masses of all pseudoscalar mesons,
vector mesons and baryons are well described in terms of naive quark model.\\
At present there is also not much known about the relative contribution of the possible 
reaction mechanisms of the production of the $\eta'$ meson. It is expected
that the $\eta'$ meson can be produced through heavy meson exchange, through the excitation
of an intermediate resonance or via emission from the virtual meson~\cite{KNakayama}.
\begin{figure}
  \includegraphics[height=.4\textheight]{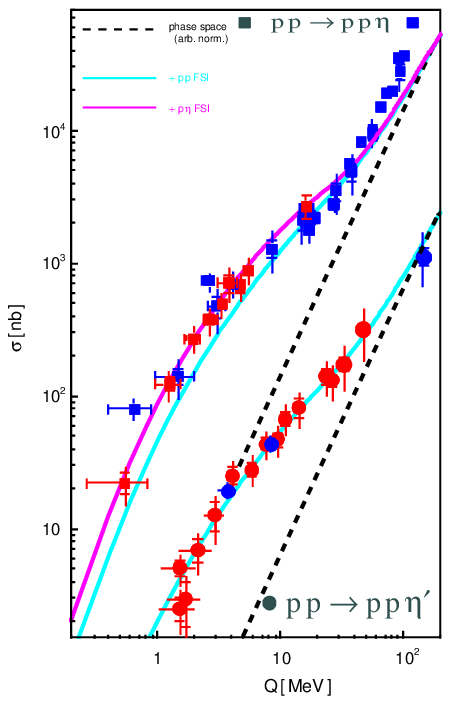}
  \caption{Total cross section for the reactions $pp \to pp\eta'$
           (circles) and $pp \to pp\eta$ (squares)as a function of the
           centre--of--mass excess energy Q. Data are from references
           \cite{PM1,AK,JS,PM2,HC,ECh,AMB,FH,FB,PM6}}
\end{figure}

However it is not possible to judge about the mechanism responsible for the $\eta'$
meson production only from the total cross section of the $pp \to pp\eta'$ reaction.
Therefore one has to investigate the $\eta'$ production in proton-neutron scattering
in order to impose the additional constraint on the existing theoretical models.
Moreover, there is also one significant aspect of these studies, viz
the close--to--threshold production of $\eta'$ meson in the nucleon--nucleon collisions 
requires a large momentum transfer between nucleons and occurs at distances in 
the order of 0.3 fm and this implies that the quark--gluon degrees of freedom may 
play a significant role in the production dynamics of this meson. Therefore it 
is possible that the $\eta'$ meson is created from excited glue in the interaction 
region of the colliding nucleons~\cite{SBass1,SBass2,SBass3}.
It is interesting to note that recently Bass and Thomas~\cite{bassthomas}
argued that the strength of the interaction  of $\eta$ and $\eta^{\prime}$ mesons
with nucleons is sensitive to the singlet-flavour component, 
and hence to the glue in these mesons. This makes a connection in our 
endeavour to investigate the structure, the production dynamics, and 
the interaction of the $\eta$ and $\eta^{\prime}$  mesons with nucleons.

\section{Isospin degrees of freedom}

Treating proton and neutron as different states of nucleon
distinguished only by the isospin projection, $+{{1}\over{2}}$ for the proton
and $-{{1}\over{2}}$ for the neutron, we may classify the $NN \to NNX$ reactions
according to the total isospin of the nucleon pair in the initial and final state.
A total isospin of two nucleons equals 1 for proton--proton and
neutron--neutron pairs, and may acquire the value 1 or 0 for the 
neutron--proton system. Since $\eta'$ meson is isoscalar, there are only two pertinent
transitions for the $NN \to NNX$ reaction, provided that it occurs via the isospin 
conserving interaction. It is enough to measure two reaction channels for 
an unambiguous determination of isospin 0 and 1 cross section~\cite{PMoskal3}.\\
As discussed in references~\cite{SBass1,SBass2,SBass3} 
the comparison of the $ pp \to pp \eta'$
and $pn \to pn \eta'$ total cross sections will allow not only to learn about the production
of the $\eta'$ meson in the channels of isospin I = 0 and I = 1 but also to investigate
aspects of the gluonium component of the $\eta'$ meson.\\
Such investigations were already performed in case of $\eta$ meson, and
the total cross section in both the proton--proton as well as the
proton--neutron reactions has been measured.
The ratio $R_{\eta} = {{\sigma({pn \to pn\eta})} \over {\sigma({pp \to pp\eta})}}$
was determined to be about 6.5~\cite{HCalen} in the excess energy range between
16 MeV and 109 MeV. Since,
$$ \sigma(pp \to pp\eta) = \sigma_{I=1},$$
$$ \sigma(pn \to pn\eta) = {{ \sigma_{I=0} + \sigma_{I=1}} \over 2}$$
we have
$$ \sigma_{I=0} = (2R_{\eta} - 1)\sigma_{I=1}, $$
where I denotes the total isospin in the initial and final state 
of the nucleon pair. This means that the production of $\eta$ meson
with the total isospin I=0 exceeds the production with the isospin I=1
by a factor of 12.
This large difference of the total cross sections suggests the 
dominance of isovector mesons exchange in the creaction of
$\eta$ in collisions of nucleons. \\
Since the quark structure of $\eta$ and $\eta'$ mesons is very similar,
in case of the dominant isovector meson exchange -- by the analogy to
the $\eta$ meson production -- we can expect that the ratio $R_{\eta'}$
should be about 6.5. If however $\eta'$ meson is produced via its
flavour-blind gluonium component from the colour-singlet glue excited
in the interaction region the ratio should approach unity after corrections
for the initial and final state interactions~\cite{SBass1}. 

\section{Status of the experiment}

The close--to--threshold
excitation function for the $pp \to pp\eta'$ reaction (see fig.1)
has been already determined,
whereas the total cross section for the $\eta'$ meson production
in proton-neutron interaction is still unknown.
Therefore we have performed the measurement of the quasi--free $pn \to pnX$ 
processes using a proton beam and a deuteron cluster target.
In two separated runs we have measured the $pn \to pnX$ process close to the 
$\eta$ and $\eta'$ production threshold. For the preliminary results on the
$pn \to pn\eta$ process the interested reader is referred to reference~\cite{PMoskal4}.
The experiment is based on the registration of all outgoing nucleons 
from the $pd\to p_{sp} pnX$ reaction. Protons are measured in two drift
chambers and scintillator detectors~\cite{Brauksiepe,Jurek}, neutrons are registered
in the neutral particle detector~\cite{przerwa1,przerwa2}.
Protons considered as a spectators are measured by the dedicated
silicon-pad detector~\cite{bilger,janusz}.
Application of the missing mass technique allows to identify events
with the creation of the meson under investigation
and the total energy available for the quasi-free proton-neutron
reaction can be calculated for each event from the vector of the momentum
of the spectator proton. For more detailed description of the experimental
method see references~\cite{PMoskal5,PKlaja,JStepaniak}.\\
At present the analysis aiming for establishing the excitation
function for the $pn \to pn\eta'$ reaction is in progress
and will deliver the values for the total cross section
in the excess energy range between 0 and 20 MeV.

\begin{theacknowledgments}
The work has been supported by the European Community - Access to
Research Infrastructure action of the Improving Human Potential Programme,
by the FFE grants (41266606 and 41266654) from the Research Centre J{\"u}lich,
by the DAAD Exchange Programme (PPP-Polen),
by the Polish State Committe for Scientific Research
(grant No. PB1060/P03/2004/26),
and by the RII3/CT/2004/506078
- Hadron Physics-Activity -N4:EtaMesonNet.
\end{theacknowledgments}


\begin{thebibliography}{9}

\bibitem{PM1}
P.~Moskal et al., \emph{Phys. Lett} {\bf B 474} (2000) 416.
\bibitem{AK}
A.~Khoukaz et al., \emph{Eur. Phys. J.} {\bf A 20} (2004) 345.
\bibitem{JS}
J.~Smyrski et al., \emph{Phys. Lett.} {\bf B 474} (2000) 182.
\bibitem{PM2}
P.~Moskal et al., \emph{Phys. Rev. Lett} {\bf 80} (1998) 3202.
\bibitem{HC}
H. Cal\'{e}n et al., \emph{Phys. Lett.} {\bf B 366} (1996) 39.
\bibitem{ECh}
E.~Chiavassa et al., \emph{Phys. Lett.} {\bf B 322} (1994)270.
\bibitem{AMB}
A.~M.~Bergdolt et al., \emph{Phys. Rev.} {\bf D 48} (1993) R2969.
\bibitem{FH}
F.~Hibou et al., \emph{Phys. Lett.} {\bf B 438} (1998) 41.
\bibitem{FB}
F.~Balestra et al., \emph{Phys. Lett.} {\bf B 491} (2000) 29.
\bibitem{PM6}
P.~Moskal et al., \emph{Phys. Rev.} {\bf C 69} (2004) 025203.
\bibitem{KNakayama}
K.~Nakayama et al., \emph{Phys. Rev.} {\bf C 61} (2000) 024001.
\bibitem{SBass1}
S. D. Bass, \emph{Phys. Lett.} {\bf B 463} (1999) 286.
\bibitem{SBass2}
S. D. Bass, e-Print Archive: hep-ph/0006348.
\bibitem{SBass3}
S. D. Bass, \emph{Phys. Scripta} {\bf T 99} (2002) 96.
\bibitem{bassthomas}
S. D. Bass, A. W. Thomas, e-Print Archive: hep-ph/0507024.
\bibitem{PMoskal3}
P.~Moskal, e-Print Archive: hep-ph/0408162.
\bibitem{HCalen}
H. Cal\'{e}n et al., \emph{Phys. Rev.} {\bf C 58} (1998) 2667.
\bibitem{PMoskal4}
P.~Moskal et al., e-Print Archive: nucl-ex/0311003.
\bibitem{Brauksiepe}
S.~Brauksiepe et al., \emph{ Nucl. Instr. $\&$ Meth.} {\bf A 376} (1996) 397.
\bibitem{Jurek}
J.~Smyrski et al., \emph{Nucl. Instr. $\&$ Meth.} {\bf A 541} (2005) 574.
\bibitem{przerwa1}
J.~Przerwa, e-Print Archive: hep-ex/0408016, \\
Berichte des FZ-J\"ulich, {\bf J\"ul-4141} (2004).
\bibitem{przerwa2}
J.~Przerwa et al., \emph{Int. J. of Mod. Phys.} {\bf A 20} (2005) 625.
\bibitem{bilger}
R.~Bilger et al., \emph{Nucl. Instr. $\&$ Meth.} {\bf A 457} (2001) 64.
\bibitem{janusz}
M.~Janusz, diploma thesis, Jagellonian University, Cracow (2004).
\bibitem{PMoskal5}
P.~Moskal, e-Print Archive: nucl-ex/0110001.
\bibitem{PKlaja}
P.~Klaja et al., e-Print Archive: hep-ex/0507055, these proceedings.
\bibitem{JStepaniak}
J.~Stepaniak, H. Cal\'{e}n, e-Print Archive: nucl-ex/0412025.

\end{thebibliography}
\end{document}